\documentclass{Interspeech2024}
\usepackage{multirow,multicol}
\usepackage{url,cite,hyperref}
\usepackage{caption,subcaption}
\usepackage{amsmath,mathrsfs}
\usepackage{graphicx}
\usepackage[subtle]{savetrees}

\renewcommand{\vec}[1]{\boldsymbol{\mathrm{#1}}}

\renewcommand{\vec}{\boldsymbol} 
\newcommand{\newpara}[1]{\vspace{2pt}\noindent\textbf{#1}}

\interspeechcameraready 
\title{Beyond Silence:\\ Bias Analysis through Loss and Asymmetric Approach in Audio Anti-Spoofing}
\name[affiliation={1}]{Hye-jin}{Shim}
\name[affiliation={2}]{Md}{Sahidullah}
\name[affiliation={1}]{Jee-weon}{Jung}
\name[affiliation={1}]{Shinji}{Watanabe}
\name[affiliation={3}]{Tomi}{Kinnunen}

\address{
  $^1$Carnegie Mellon University, USA\\
  $^2$TCG CREST, India \\
  $^3$University of Eastern Finland, Finland}
\email{shimhz6.6@gmail.com
}

\begin{document}

\maketitle

\begin{abstract}
Current trends in audio anti-spoofing detection research strive to improve models' ability to generalize across unseen attacks by learning to identify a variety of spoofing artifacts. This emphasis has primarily focused on the spoof class.
Recently, several studies have noted that the distribution of silence differs between the two classes, which can serve as a shortcut. In this paper, we extend class-wise interpretations beyond silence. We employ loss analysis and asymmetric methodologies to move away from traditional attack-focused and result-oriented evaluations towards a deeper examination of model behaviors. Our investigations highlight the significant differences in training dynamics between the two classes, emphasizing the need for future research to focus on robust modeling of the bonafide class.
\end{abstract}

\section{Introduction}
\vspace{-0.2cm}
Recent progress in voice conversion (VC) and text-to-speech (TTS) technologies have intensified concerns regarding their potential for malicious use, emphasizing the critical role of audio anti-spoofing systems.
Audio spoofing detection systems, as binary classifiers, distinguish genuine human speech ({\em bonafide}) from artificially generated ({\em spoofed}) speech.
The main direction in this field is toward advancing the systems' ability to generalize across unseen spoofing attacks by focusing on learning diverse spoofing artifacts.
To this end, a pivotal shift from traditional hand-crafted features~\cite{sahidullah2015comparison, todisco2017constant} to data-driven approaches~\cite{tak2021rawnet2, jung2019replay, tak2021end} and data augmentation~\cite{das2021data, tak2022rawboost} have been pursued.
Research on diversifying training data~\cite{das2020assessing, paul2017generalization, shim2023multi}, incorporating domain adaptation strategies~\cite{himawan2019deep,xie2023domain}, and leveraging large-scale pre-trained models~\cite{wang2022investigating, tak2022automatic, wang2023can} also follow this trend.

Despite these efforts, the evolution of TTS and VC systems, especially those utilizing state-of-the-art methods like diffusion, are likely to involve fewer spoofing artifacts and create subtle variations in them without hard effort.
This suggests that learning diverse spoofing artifacts does not guarantee the detection of emerging unknown attacks.
To tackle this, several studies have explored methods for modeling robust bonafide features and distinguishing them from spoof features within the latent space~\cite{chen2020generalization, zhang2021one, zhang2023impact, ren2023lightweight}.
Class-wise analyses have sought to determine the differences between bonafide and spoof classes as detailed in \cite{chettri2021data, liu2023asvspoof, zhang2023impact} and their findings primarily converge on the impact of `silence', which has a spurious correlation with spoofing detection.

\begin{figure}[t!]
    \includegraphics[width=\linewidth]{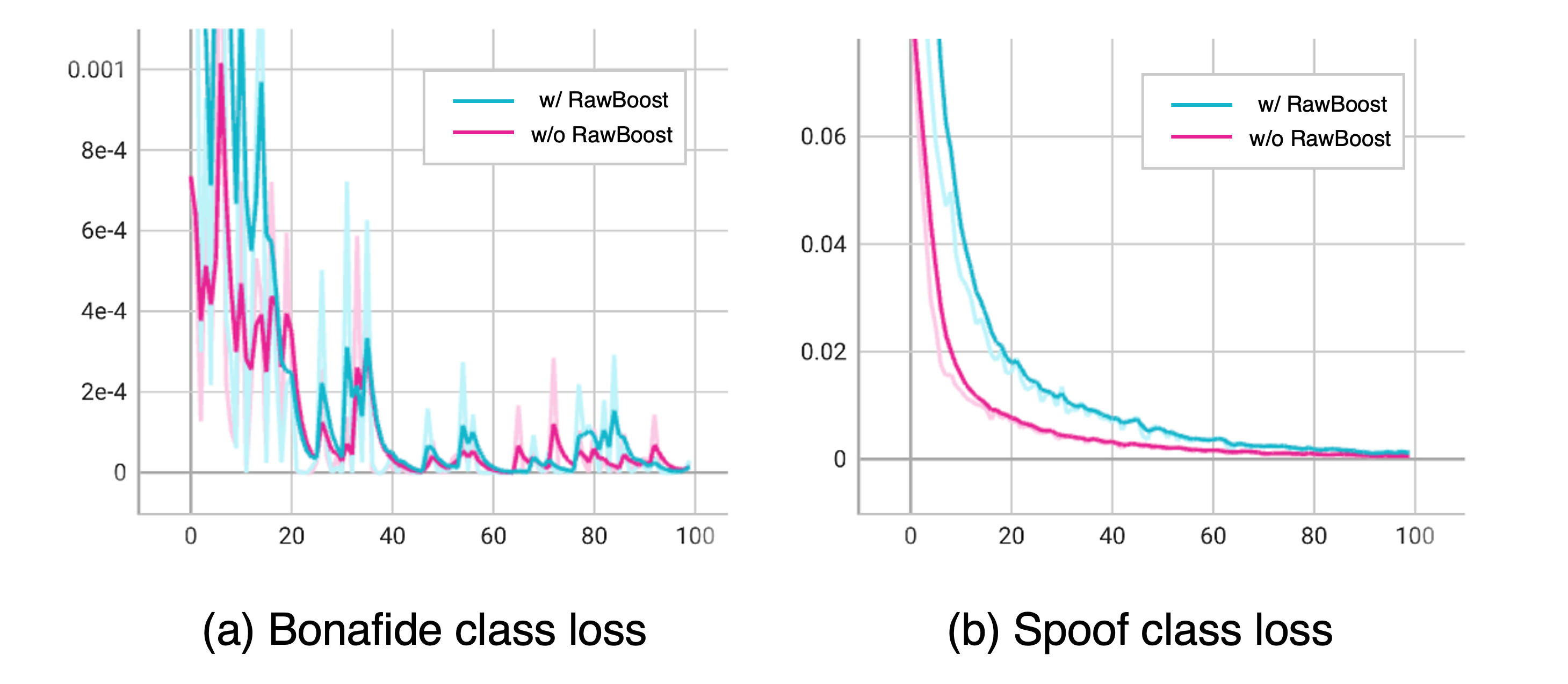}
    \vspace{-0.8cm}
\caption{Comparison of training loss. The left and right figures illustrate bonafide and spoof classes. x-axis and y-axis indicate training epochs and loss magnitude. Regardless of the implementation of data augmentation, the two class losses differ on a large scale.}
\label{fig:loss_trend1}
\vspace{-0.7cm}
\end{figure}

These kinds of external factors that could unintentionally lead the biased model predictions are known as shortcuts~\cite{geirhos2020shortcut}, recently getting a lot of attention in the broader deep learning literature.
These shortcuts challenge the ability to determine if a model genuinely distinguishes between classes or relies on irrelevant cues associated with class labels.
In the context of audio anti-spoofing, `silence' is a well-known \textit{data bias}; for instance, the model trained with ASVspoof2017 dataset~\cite{kinnunen2017asvspoof} is vulnerable to silence~\cite{chettri2020dataset} and the ASVspoof2019 dataset~\cite{todisco2019asvspoof} exhibits an unequal distribution of silence between spoofed and bonafide samples~\cite{muller21_asvspoof, zhang2021effect}. 
The impact of silence varies not only with the data source but also with its distinct types/patterns introduced during data collection.
Efforts to understand and mitigate the influence of silence, including the analysis of silence trimming and its effects, have been undertaken~\cite{muller21_asvspoof, zhang2023impact}.
Other than silence, unveiled factors also have been studied in~\cite{Hyejin2023-coin-flip}, recently.

While in-depth analyses in audio anti-spoofing have significantly advanced the field and provided invaluable insights, several areas remain for further explorations: (i) the majority of studies have focused on spoof class, specifically per-attack interpretations; (ii) analyses on each class exist, yet the emphasis has primarily been on silence, except for \cite{Hyejin2023-coin-flip} exploring other data bias factors; 
(iii) existing analyses predominantly rely on evaluating outcomes, such as performance metrics and score distributions, without a thorough examination of the internal workings of models during training.
Our research diverges from existing studies by delving into the training process through loss analysis and adopting a novel asymmetric intervention on each class and phase(train and test) to understand the respective effects.

Our analysis starts with examining the loss curves for bonafide and spoof classes separately, both with and without the application of RawBoost~\cite{tak2022rawboost} data augmentation, as illustrated in \autoref{fig:loss_trend1}.
Note that illustrated loss values are raw values. For the model update, a loss weight of 0.9 (bonafide) and 0.1 (spoof) was employed, considering the imbalanced number of samples in each class. 
Interestingly, the results indicate that the loss associated with the bonafide class is significantly lower than that for the spoof class. Even if one considers the loss weights in the training phase, a substantial gap still remains between the two classes. 
These results could signify that the bonafide class is inherently easier to train than the spoof class. 
Alternatively, this might suggest that modeling the bona fide class is not trivial; however, there exists a shortcut that significantly reduces the training loss.
We deploy various loss functions to reveal the meaning behind the low magnitude of the bonafide loss, between the two scenarios mentioned.
In particular, the objective functions that can consider the difficulty of samples enable us to understand how the model deals with each class based on our findings.

Furthermore, our approach involves training and testing the model with an asymmetric intervention to assess the bonafide and spoof classes separately.
Unlike the methodology in \cite{Hyejin2023-coin-flip}, which applies interventions across both classes and phases for a comprehensive model-level interpretation, our strategy focuses interventions on one side only.
This allows a more precise understanding of how different phases and class-specific traits affect model performance.
Additionally, to mitigate potential concerns regarding the influence of silence, we also conduct our analysis combined with silence trimming as well, demonstrating that our findings are not biased by the presence of silence. 
Our findings pave the way for new directions of future research, shifting focus from the currently predominant studies centered around the spoof class.

\section{Method}
\vspace{-0.2cm}
In this section, we introduce two primary methodologies that facilitate our in-depth investigations of anti-spoofing systems: (i) a loss-based analysis that contrasts objective functions, distinguishing between those that prioritize either easier or more challenging samples, and (ii) an asymmetric intervention analysis that examines the impact of intervention on phases (train or test) and classes (bonafide or spoof). 

\subsection{Loss based analysis}
\vspace{-0.2cm}
Previous analyses of silence within audio anti-spoofing contexts have paved the way for separate examinations of bonafide and spoof classes.
This study extends these insights by exploring the various objective functions that can help us understand the model's behavior in each class;
loss analysis is widely adopted in machine learning for understanding the model behavior during the training~\cite{wu2017towards}. 
Furthermore, the loss has been directly employed in the image domain to infer data bias in~\cite{nam2020learning, lee2021learning, hwang2022selecmix} by analyzing its correlation with sample difficulty and data bias.

In this work, we conduct two types of analysis using loss functions.
Initially, we calculate the loss for bonafide and spoof classes separately during training to observe their convergence patterns and magnitude differences. Significant differences in these areas may indicate model bias, as shown in~\autoref{fig:loss_trend1}. 
Subsequently, we differentiate between two categories of loss functions for assessing model behavior: prioritizing hard \textit{or} easy samples.
Among diverse loss functions utilized in hard negative mining, we deliberately choose a few that dynamically modulate each sample's impact on the overall loss, rather than explicitly selecting samples. This strategy enables an in-depth exploration of the model's inherent reactions to varying sample types.

\newpara{FocalLoss} \cite{lin2017focal} prioritizes hard samples by modifying the categorical cross-entropy loss. 
It amplifies the loss for samples with inaccurate model predictions or where the model exhibits uncertainty (i.e., low predicted probability for the correct class), thus prioritizing hard or incorrectly classified samples without resorting to specific sample selection strategies.

\newpara{SuperLoss} \cite{castells2020superloss}  assigns a weight to each sample based on a moving average of its past losses, targeting potentially noisy or outlier samples. This method progressively focuses on challenging or ambiguous samples.

\newpara{CurricularFace} \cite{huang2020curricularface} emphasizes more challenging samples by increasing their relative loss compared to easier ones using class-specific margins.
It adjusts the target margins for classes, making it easier or harder for the model to classify them correctly as training progresses.

\newpara{Generalized cross entropy (GCE)} \cite{zhang2018generalized}, in contrast, focuses on {\em easier} samples. It enhances the penalty for misclassifying classes with lower probabilities, thereby prioritizing minority classes. This method can be particularly useful in audio anti-spoofing to direct the model's attention towards bonafide samples.
In addition, GCE is employed in model debiasing research to concentrate on biased samples, aligning with empirical observations that such samples exhibit lower loss values during training.

\vspace{-0.2cm}
\subsection{Asymmetric intervention analysis}
\vspace{-0.2cm}
Our analysis follows an interventional approach proposed recently in~\cite{Hyejin2023-coin-flip} to reveal shortcuts in speech anti-spoofing beyond silence. The essence of this approach is to intentionally modify (intervene) an existing dataset to provoke the classifier to rely on shortcuts. We start by reviewing the approach and then explain how we modify it in this study.

To explore potential shortcuts in audio anti-spoofing, \cite{Hyejin2023-coin-flip} employed standard audio manipulations (e.g., MP3 compression or additive noise) with randomized parameters applied to datasets to create artificial statistical associations between the audio and the class label. 
For example, applying MP3 compression to only bonafide data in both training and test data while leaving spoof data unaltered resulted in equal error rates (EERs) of 0\%. Conversely, applying MP3 compression exclusively to spoofed test data while keeping all other conditions the same led to an opposite outcome (EER $>99\%$), illustrating a complete label flip.
Such extreme interventions reveal the vulnerability of the spoofing detection system against potential bias factors unrelated to spoofing artifacts residing within the data.

\begin{table}[!t] 
    \caption{Five experimental configurations using eight subsets of a dataset. \text{\textbf{O}} is the original (unintervened) configuration. All other four sets have one particular subset intervened.}
    \vspace{-0.2cm}
    \label{tab:config3}
    \centering
    \resizebox{\columnwidth}{!}{
    \begin{tabular}{c|c|c|c}
    \toprule
Intervened phase & Configuration & Train set & Test set \\ \hline
- & \textbf{O} & $\mathcal{D}_\text{trn, bona} \cup \mathcal{D}_\text{trn,spf}$ & $\mathcal{D}_\text{test, bona} \cup \mathcal{D}_\text{test, spf}$ \\ \hline
\multirow{2}{*}{Train} & \textbf{Tr}\_\textbf{B} & $\mathcal{D}^{\textbf{intervened}}_\text{trn, bona} \cup \mathcal{D}_\text{trn,spf}$ & $\mathcal{D}_\text{test, bona} \cup \mathcal{D}_\text{test, spf}$ \\ 
 & \textbf{Tr}\_\textbf{S} & $\mathcal{D}_\text{trn, bona} \cup \mathcal{D}^{\textbf{intervened}}_\text{trn,spf}$ & $\mathcal{D}_\text{test, bona} \cup \mathcal{D}_\text{test, spf}$ \\ \hline
\multirow{2}{*}{Test} & \textbf{Te}\_\textbf{B} & $\mathcal{D}_\text{trn, bona} \cup \mathcal{D}_\text{trn,spf}$ & $\mathcal{D}^{\textbf{intervened}}_\text{test, bona} \cup \mathcal{D}_\text{test, spf}$ \\ 
 & \textbf{Te}\_\textbf{S} & $\mathcal{D}_\text{trn, bona} \cup \mathcal{D}_\text{trn,spf}$ & $\mathcal{D}_\text{test, bona} \cup \mathcal{D}^{\textbf{intervened}}_\text{test, spf}$ \\ 
\bottomrule
    \end{tabular}
    }
    \vspace{-0.5cm}
\end{table}

The previous approach concurrently applied interventions during both training and test phases to understand the overall model's behaviors under different interventions, producing the boundary cases of `near-perfect' outcomes and `worse than coin flip' (label flip) results.
In contrast, our investigation takes a class-wise interpretive approach based on the understanding that each class responds differently. 
Consequently, we strategically apply interventions at only one phase or one class at a time, leveraging two binary dimensions: phase (training or testing) and class (bonafide or spoof). 
Our approach enables (i) to reveal the interventions associated with each class separately, (ii) to evaluate the robustness of the model's representation for each class separately, and (iii) to compare the robustness of class modeling by observing the outcomes of intervention.

In particular, our methodology of employing class- and phase-wise intervention in either class in either phase results in four distinct intervention configurations. 
Formally, let $\mathcal{D}=\{(x_i, y_i) | x_i \in \mathbb{X}, y_i \in \mathbb{Y}, \text{for}\ i=1, 2, ... ,N$,
where $\mathcal{D}$ represents the dataset, ($x_i, y_i$) denotes an individual sample, and $N$ is the number of samples.
Typically, the corpus is divided into two subsets: the training ($\mathcal{D}_\text{trn}$) and the test ($\mathcal{D}_\text{test}$) sets. 
We introduce another categorization based on class, dividing the dataset into four subsets: $\mathcal{D}_\text{trn, bona}$, $\mathcal{D}_\text{trn, spf}$, $\mathcal{D}_\text{test, bona}$, $\mathcal{D}_\text{test, spf}$. Furthermore, we generate four additional subsets with applied interventions, where the labels for each sample remain: $\mathcal{D}^{\textbf{intervened}}_\text{trn, bona}$, $\mathcal{D}^{\textbf{intervened}}_\text{trn, spf}$, $\mathcal{D}^{\textbf{intervened}}_\text{test, bona}$, $\mathcal{D}^{\textbf{intervened}}_\text{test, spf}$. Note that the intervention does not affect the ground-truth label $\mathbb{Y}$.
Composing the eight total subsets, we present five experimental configurations in Table 1: $\textbf{O}$, $\textbf{Tr}\_\textbf{B}$, $\textbf{Tr}\_\textbf{S}$, $\textbf{Te}\_\textbf{B}$, and $\textbf{Te}\_\textbf{S}$.
Therefore, the comparison of $\textbf{B}$ and $\textbf{S}$, modification on the class side, enables us to analyze the class-wise effect of interventions. 
Comparison of $\textbf{Tr}$ and $\textbf{Te}$, on the other hand, helps to understand the class-wise difference and how the model operates differently depending on whether the model learns such intervened condition.

\section{Experimental setup}
\subsection{Datasets}
\vspace{-0.2cm}
\newpara{ASVspoof2019}~\cite{todisco2019asvspoof} is a dataset for logical access (LA) scenario of audio anti-spoofing and it includes 6 and 13 types of spoofing attacks in train/dev and test.
The number of utterances for train/dev and test is $50,224$ and $71,237$, respectively.

\begin{table}[t!]
  \caption{Performance comparison on different loss functions. Apart from original categorical cross entropy (CCE), FocalLoss, SuperLoss, and CurricularFace are designed to concentrate on hard samples based on empirical loss while generalized cross entropy (GCE) focuses on easier samples. All evaluations are conducted on the ASVspoof2019 LA dataset.}
  \centering
  \vspace{-5pt}
  \label{tab:loss_compare}
  \resizebox{0.6\columnwidth}{!}{
  \begin{tabular}{lcc}
  \toprule
  Loss   function &  & EER   (\%) \\
  \hline
  Original (CCE) & &1.39 \\
  FocalLoss\cite{lin2017focal} &  &1.67 \\
  SuperLoss\cite{castells2020superloss} &  & 1.45 \\
  CurricularFace\cite{huang2020curricularface} &  & 1.53\\
  \textbf{GCE}\cite{zhang2018generalized} &  &\textbf{1.35} \\
  \bottomrule
\end{tabular}
}
\vspace{-0.2cm}
\end{table}

\newpara{ASVspoof2021}~\cite{ASV2021challenge} includes the latest spoofing attacks,
which consists of different test sets, each for LA and deepfake (DF) scenarios. 
The LA and DF subsets include diverse synthetic techniques and audio compressions, respectively.

\subsection{Implementation details}
\vspace{-0.2cm}
\newpara{Model architectures} used in this paper 
are AASIST, AASIST-L~\cite{jung2022aasist} 
 models with data augmentation and multi-dataset co-training.
Those two models are state-of-the-art models that directly operate on raw waveform
and they only differ in the number of parameters. 

\begin{table}[t!]
    \caption{Performance comparison in EER when the model is \textbf{tested} on asymmetric ways. All models are trained on the original training dataset but evaluated in different configurations. Here, the ratio indicates a relative change of \textbf{Te\_B} compared to that of \textbf{Te\_S}.} 
    \centering
    \label{tab:test_interv}
    \vspace{-2pt}
    \resizebox{0.9\columnwidth}{!}{
    \begin{tabular}{lcccc}
    \toprule
    System & \textbf{O} & \textbf{Te\_S} & \textbf{Te\_B} & Ratio \\
    \hline \hline
    \textbf{MP3} & & & & \\
    \hline
    AASIST & 0.83 & 0.77 & 9.07 & 137.33 \\
    AASIST-L & 1.39 & 1.21 & 8.78 & 41.06 \\
    AASIST-L   w/ RawBoost & 1.59 & 1.62 & 3.96 & 79.00 \\
    AASIST-L   w/ MDL & 0.99 & 1.29 & 3.63 & 8.80 \\
    \hline
    \textbf{Noise} & & & & \\
    \hline
    AASIST & 0.83 & 0.39 & 31.08 & 68.75\\
    AASIST-L & 1.39 & 0.65 & 35.50 & 46.09\\
    AASIST-L   w/ RawBoost & 1.59 & 1.06 & 13.27 & 22.04 \\
    AASIST-L   w/ MDL & 0.99 & 0.69 & 8.24 & 24.17 \\
    \hline
    \textbf{Loudness} & & & & \\
    \hline
    AASIST & 0.83 & 1.18 & 5.33 & 12.86\\
    AASIST-L & 1.39 & 2.23 & 6.73 & 6.36\\
    AASIST-L   w/ RawBoost & 1.59 & 1.52 & 2.38 &  11.29\\
    AASIST-L   w/ MDL & 0.99 & 1.44 & 4.09 & 6.89 \\
    \bottomrule
\end{tabular}
}
\vspace{-0.2cm}
\end{table}

\newpara{Intervention types} ~are selected among five different interventions in~\cite{Hyejin2023-coin-flip}. We employ three interventions: \emph{MP3 compression}, \emph{additive white noise}, and \emph{loudness normalization}. Those interventions are considered since MP3 compression and white noise are discovered as the most influential ones, while loudness normalization is the least effective one. 

\newpara{Data augmentation}~is implemented to check the difference when we employ the model considered more robust. We utilize RawBoost~\cite{tak2022rawboost} which includes three different augmentation techniques: linear and non-linear convolutive noise, multi-band filters, and Hammerstein systems~\cite{kibangou2006wiener}.
We deploy three of them simultaneously as it showed the best result in ~\cite{tak2022rawboost}.

\begin{table*}[t]
  \caption{Performance comparison in EER when the model is \textbf{tested} on asymmetric ways with silence trimming. Silence trimming is applied in both the training and the test phases; we additionally report the results when the silence is only removed from either phase. 
  } 
  \vspace{-5pt}
  \centering
  \label{tab:silence_trim}
  \resizebox{\textwidth}{!}{
    \begin{tabular}{l|l|c|ccc|ccc|ccc}
    \toprule
    \multicolumn{1}{c|}{\multirow{2}{*}{System}} & \multicolumn{1}{c|}{\multirow{2}{*}{Silence   trimming}} & \multicolumn{1}{c|}{\multirow{2}{*}{\textbf{O}}} & \multicolumn{3}{c|}{MP3} & \multicolumn{3}{c|}{Noise} & \multicolumn{3}{c}{Loudness} \\
    \multicolumn{1}{c|}{} & \multicolumn{1}{c|}{} & \multicolumn{1}{c|}{}& \textbf{Te\_S} & \textbf{Te\_B} & Ratio & \textbf{Te\_S} & \textbf{Te\_B} & Ratio & \textbf{Te\_S} & \textbf{Te\_B} & Ratio \\
    \hline
    AASIST-L & - & 1.39 & 1.21 & 8.78 & 41.06 & 0.65 & 35.50 & 46.09 & 2.23 & 6.73 & 6.36 \\
    AASIST-L w/ RawBoost & - & 1.59 & 1.62 & 3.96 & 79.00 & 1.06 & 13.27 & 22.04 & 1.52 & 2.38 & 11.29 \\ \hline \hline
    AASIST-L & train & 35.05 & 28.36 & 39.32 & 0.64 & 47.09 & 28.14  & 0.57 & 27.44 &  47.28 & 1.61 \\
     & test & 25.14 & 20.07 & 37.7 & 2.48 & 8.27 & 59.04 & 2.01 & 13.92 & 40.83 & 1.40 \\
     & train\&test & 18.65 & 17.52 & 30.65 & 10.62 & 17.24 & 34.49 & 11.23 & 14.79 & 26.99 & 2.16 \\ \hline
    
    AASIST-L w/ RawBoost & train & 45.52 & 40.65 & 48.47 & 0.61 &  66.93 & 31.14 & 0.47 &36.10 & 55.35 & 1.04\\
     & test  & 27.79 & 29.5 & 30 & 1.29 & 30.74 & 26.85 & 0.32 & 21.82 & 37.05 & 1.55 \\
     & train\&test & 19.73 & 17.56 & 31.69 & 5.51 & 21.07 & 32.35 & 9.42 & 12.75 & 29.73 & 1.43 \\
    \bottomrule
    \end{tabular}
}
    \vspace{-0.5cm}
\end{table*}
\newpara{Multi-dataset trained model with sharpness optimization} ~\cite{shim2023multi} is utilized to investigate the robust model similar to data augmentation.  To further the enhance generalization capability, the model is trained using multiple datasets at the same time and optimized by sharpness-aware optimization~\cite{foret2020sharpness}. 
We select the model trained by both ASVspoof2015 and ASVspoof2019 LA with adaptive sharpness-aware minimization (ASAM)~\cite{kwon2021asam} as it showed the best performance in ASVspoof2019 LA evaluation by 0.99\% of EER.

\newpara{Silence trimming} is implemented 
to mitigate the influence of silence that might distort the imbalance results. Our silence trimming algorithm works as follows. First, we detect the speech frames using a simple energy-based algorithm as described in~\cite{kinnunen2010overview}. We have used a frame size of $25$~ms with an $8$ms shift. Then we remove the silences where the silence length is more than $50$~ms.


\begin{table}[t!]
\centering
\caption{Performance comparison in EER when the model \textbf{trained} on asymmetric ways. All evaluations are conducted using original test set without interventions.}
\label{tab:train_interv}
\vspace{-5pt}
\resizebox{\columnwidth}{!}{
\begin{tabular}{lccccc} 
\toprule
\multicolumn{1}{c}{System} & \begin{tabular}[c]{@{}c@{}}Intervention\\ type\end{tabular} & Config.        & \begin{tabular}[c]{@{}c@{}}2019 \\ LA\end{tabular} & \begin{tabular}[c]{@{}c@{}}2021 \\ LA\end{tabular} & \begin{tabular}[c]{@{}c@{}}2021\\ DF\end{tabular}  \\ 
\hline
Original                   & -                                                           & -              & 1.39                                               & 12.18                                              & 21.8                                               \\ 
\hline
RawBoost                   & -                                                           & -              & 1.59                                               & 6.42                                               & 17.48                                              \\ 
\hline\hline
\multirow{6}{*}{RawBoost}  & \multirow{2}{*}{MP3}                                        & \textbf{Tr\_B} & 16.21                                              & 15.45                                              & 30.86                                              \\
                           &                                                             & \textbf{Tr\_S} & 3.68                                               & 17.81                                     & 23.1                                               \\ 
\cline{2-6}
                           & \multirow{2}{*}{Noise}                                      & \textbf{Tr\_B} & 71.02                                              & 62.45                                              & 62.66                                              \\
                           &                                                             & \textbf{Tr\_S} & 15.48                                              & 24.12                                              & 30.06                                              \\ 
\cline{2-6}
                           & \multirow{2}{*}{Loudness}                                   & \textbf{Tr\_B} & 13.05                                              & 10.66                                              & 18.7                                               \\
                           &                                                             & \textbf{Tr\_S} & 1.82                                               & 9.49                                               & 17.97                                              \\
\bottomrule
\end{tabular}
}
\vspace{-0.6cm}
\end{table}
\section{Results and Analysis}

\subsection{Comparison of class-wise loss and different objective functions}
\vspace{-0.2cm}
As previously introduced in Section 1 with~\autoref{fig:loss_trend1}, the observed loss curve reveals an unexpected pattern, especially considering the predominance of spoof data in the training set; in general, more frequently represented spoof classes are expected to have lower losses, consistent with the general bias of neural networks to focus on more frequently represented classes. 
The results indicate that effective training of the bonafide class is hindered not only by its fewer amount of samples in the training dataset but also by its comparatively lower loss. Consequently, the neural network would prioritize minimizing the spoof class's loss in which the loss scale is much higher, leading to an inherent bias, which is unintended. 
This research marks the first to identify this particular bias within the context of audio spoofing detection, shedding light on the challenges posed by class imbalance and its impact on model training dynamics.

Building upon these insights, we further examine the potential class bias in model training towards the spoof class using diverse loss functions outlined in Section 2.1. 
Our comparison, shown in \autoref{tab:loss_compare}, involves four different loss functions -- three (FocalLoss~\cite{lin2017focal}, SuperLoss~\cite{castells2020superloss}, and CurricularFace~\cite{huang2020curricularface}) designed for challenging (spoof) samples and one (GCE\cite{zhang2018generalized} for easier (bonafide) samples.
The results reveal a decline in performance with the three loss functions aimed at the spoof class, while the GCE, which prefers the bonafide class, slightly improves model performance. Despite adjustments using tunable parameters in FocalLoss, SuperLoss, and CurricularFace, all outcomes showed no significant enhancement\footnote{Presented performances are the best results among our experiments.}.
Results highlight the training bias towards the spoof class.
This pattern of bias and its impact on performance encourages us to call for a significant shift in focus for future audio anti-spoofing efforts, emphasizing \textbf{the importance of robust modeling the bonafide class} over solely concentrating on detecting spoofing artifacts. 
In addition, the field of anomaly sound detection~\cite{chandola2009anomaly} further demonstrates the advantages of focusing on bonafide modeling. State-of-the-art systems prioritize modeling the normal sound class and effectively identify deviations significantly distant from these norms as anomalies. 
\subsection{Asymmetric test results}
\vspace{-0.2cm}
\autoref{tab:test_interv} presents the results of applying asymmetric interventions to a specific class during the test phase. 
These results are analyzed from two angles: (i) evaluating the intervention's impact by comparing the original results (\textbf{O}) with those after interventions on the spoof (\textbf{Te}\_\textbf{S}) or bonafide (\textbf{Te}\_\textbf{B}) classes (columns 2 and 3), and (ii) examining how interventions differently affect the bonafide class compared to the spoof class, as indicated in column 4. 
From the first angle, enhancements in performance post-intervention hint at possible model biases due to shortcuts. Conversely, declines indicate the intervention's irrelevance to the target class or might reflect domain differences or unknown factors. 
The ratio in column 4, calculated as the relative performance impact $(|\textbf{O} - \textbf{Te}\_\textbf{B}| / \textbf{O}) / (|\textbf{O} - \textbf{Te}\_\textbf{S}| / \textbf{O})$, offers deeper insight into which class the intervention impacts more.
Here, $\text{ratio} > 1$ refers to a greater influence on the bonafide class, while $\text{ratio} < 1$ implies spoof class is more affected. 

Our findings highlight two main observations that align with our analysis of loss functions. First, interventions are prone to improve \textbf{Te}\_\textbf{S} (e.g., AASIST with MP3 or AASIST-L with Noise on \textbf{Te}\_\textbf{S}), while consistently lowering \textbf{Te}\_\textbf{B} performance. This implies that factors leading to quality degradation in utterances tend to be linked with the spoof class, helping the model to classify such inputs as spoofed. The drop in \textbf{Te}\_\textbf{B} performance may result from interventions causing bonafide utterances to resemble spoofed ones more closely. Second, the ratio always surpasses 1, as a universal trend regardless of which intervention applied. This emphasizes the bonafide class's modeling fragility, suggesting that future research should more focus on the robust modeling of the bonafide class.

\subsection{Further test interventions with silence trimming}
\vspace{-0.2cm}
As introduced in Section 1, the silence could significantly influence the class-wise imbalance results. 
This prompts an important question: ``Do our observations remain valid when silence is excluded from the training or testing datasets or both?"
To address this, we conducted experiments with silence removed -- referred to as silence trimming -- to assess its effect on our findings and show the results in ~\autoref{tab:silence_trim}.
The results show a remarkable consistency across most of our results.
When silence was eliminated from both phases, effectively eliminating its influence, the ratio consistently exceeded 1, reaching a peak of 11.23.
In additional experiments where silence was removed from only one phase, there were a few instances where the ratio fell below 1; however, the majority still exhibited ratios significantly greater than 1. These consistent outcomes robustly support the conclusion that the presence of silence does not skew our findings.

\subsection{Asymmetric training results}
\vspace{-0.2cm}
Lastly, we focus on asymmetric interventions during the training phase to explore two key questions: (i) ``What happens when interventions are applied solely during training?" and (ii) ``Does the class-specific effect of interventions remain consistent if the test set remains untouched?"\footnote{The second question is related to the analysis in Section 4.2, where we want to clarify whether performance degradation stems from suddenly appearing unseen domains in the test phase.}
According to the findings presented in \autoref{tab:train_interv}, an improvement was observed in only one out of nine instances for the \textbf{Tr}\_\textbf{S} condition. 
In contrast, interventions generally led to diminished performance in the bonafide class, highlighting the necessity for more advanced methods to accurately model genuine speech.
Please note that we aim to compare the value between \textbf{Tr}\_\textbf{B} and \textbf{Tr}\_\textbf{S} for each condition to understand how each class is affected by each intervention, not the improvement from the baselines in the first two rows. 
Our findings diverge from those in \cite{Hyejin2023-coin-flip}, with EER deteriorating under all asymmetric intervention conditions except for one.
This discrepancy is likely due to our one-side distinct strategy of applying interventions exclusively during either the training or testing phase, creating a mismatch. 
Nevertheless, similar to findings in \cite{Hyejin2023-coin-flip}, we also observed that interventions involving loudness were generally less impactful compared to other types of interventions across all evaluations.

\section{Conclusion}
\vspace{-0.2cm}
This paper has conducted an in-depth investigation into the behavior of audio anti-spoofing models through various experiments focused on loss analysis and asymmetric interventions. Our analyses expand the perspective beyond attack-centric or silence-focused interpretations. The findings suggest that current training practices, which primarily aim to detect spoofing artifacts in known attacks, may neglect the robust modeling of bona fide speech, potentially introducing bias in model learning.
By advocating for a more balanced focus on understanding both bona fide and spoofed classes, our research paves the way for future studies to enhance the efficacy of audio anti-spoofing systems.

\clearpage
\section{Acknowledgment}
\vspace{-0.2cm}
The work has been partially supported by the Academy of Finland (Decision No. 349605, project “SPEECHFAKES”). Experiments of this work used the Bridges2 system at PSC and Delta system at NCSA through allocations CIS210014 and IRI120008P from the Advanced Cyberinfrastructure Coordination Ecosystem: Services \& Support (ACCESS) program, supported by National Science Foundation grants \#2138259,\#2138286, \#2138307, \#2137603, \#2138296.

\bibliographystyle{IEEEtran}
\bibliography{shortstrings, mybib}

\end{document}